\documentclass[10pt,letterpaper]{article}
\usepackage{opex3}
\usepackage{amsmath}
\usepackage{cite}

\begin{document}

\title{Dispersion Properties of Multilayered Metal-Dieletric Metamaterials}

\author{Alexey A. Orlov$^1$, Ivan V. Iorsh$^{1,2}$, Pavel A. Belov$^{1}$, Yuri S. Kivshar$^{1,3}$}

\address{$^1$ National Research University of Information Technologies, Mechanics and Optics \\ 197101, Kronverksky pr. 49, St. Petersburg, Russian Federation \\
$^{2}$St. Petersburg Academic University - Nanotechnology Research and Education Centre, St.~Petersburg 194021, Russia Federation \\
$^{3}$Nonlinear Physics Centre, Research School of Physics and Engineering, Australian National University, Canberra ACT 0200, Australia
}

\email{alexey.orlov@phoi.ifmo.ru} 


\begin{abstract}
We study dispersion properties of layered metal-dielectric media having different layers thicknesses ratios. Plotting dispersion diagrams and isofrequency contours, we find that strong nonlocalality is inherent property of such periodic structures. We introduce different level of losses and analyze complex modes of the metamaterial demonstrating that it operates in a regime with infinite numbers of eigenmodes, with nonlocality slightly affected by losses. 
\end{abstract}


\bibliographystyle{osajnl}
\bibliography{oe_orlov}

\section{Introduction}

Multilayered structures are well-known basic structures in optics. They have simple geometry, therefore such structures can be studied with ease. Nevertheless, this simplicity hasn't prevented multilayered structures from catching a new wave of attention last years. Optical metamaterials formed by multilayered metal-dielectric nanostructures (MMDN) were found to have striking electromagnetic properties. An impressive example is the fact that the structure, being treated as an indefinite medium~\cite{smith_schurig}, is able to experience effective metric signature changes from Minkowski-like $(- + + +)$ to 2T physics $(- - + +)$~\cite{smolyaninov_prl}.  A number of applications exploiting unique properties of the multilayered metamaterial were proposed, namely subwavelength imaging~\cite{shamonina,Pendry-Ramakrishna,Belov,Wood-Pendry,scalora2,he,Salandrino,Narimanov,HyperScience}, nanolithography~\cite{xiong}, optical nanocircuits~\cite{metactronics}, 3D negative refraction~\cite{3drefr}, and invisibility~\cite{cloaking}. Recently, multilayered metamaterials have become widely considered as a realization of the so-called hyperbolic media~\cite{lindell_2001} with resulting ultra-high values for the Purcell factor~\cite{sasha_pra,vanya_2012}.

\begin{figure}
\centering\includegraphics[width=0.6\textwidth]{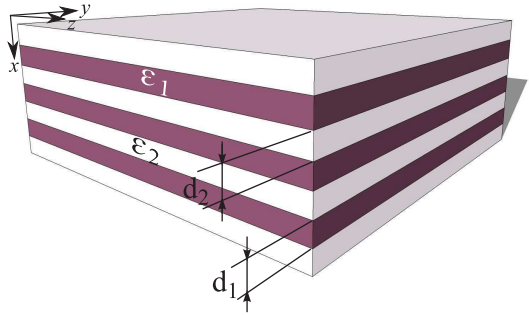}
\caption{Scheme of the structure. \label{fig:structure}}
\end{figure}

It is known that the structure under consideration demonstrates a strong nonlocal response~\cite{Podolskiynonlocal}. Previously, we have shown in~\cite{orlovPRB} that under certain conditions two extraordinary waves instead of one can be excited in the metamaterial by incident TM-polarized light beam. The existence of such an additional light wave is a clear manifestation of the presence of spatial dispersion in the metamaterial, and it validates strong nonlocal nature of the structure. In the recent paper~\cite{alu-nonlocal} this phenomenon was treated in the framework of arising field of nonlocal transformation optics. An important question of will the nonlocality and the beam splitting phenomenon still remain in real structures with introduction of losses emerges though. Consequently, it is needed to study complex modes of the multilayered metamaterial and clarify an effect of losses.  

Local effective medium model along with transfer matrix method~\cite{born-wolf} are used in order to describe the structure. The former models the structure depicted in Fig.~\ref{fig:structure} with layers permittivities $\varepsilon_1, \varepsilon_2$ and thicknesses $d_1, d_2$ as the uniaxial anisotropic crystal of the following form:
\begin{equation}
\varepsilon_\mathrm{eff} = \left( \begin{matrix} \varepsilon_\bot & 0 & 0\\0 & \varepsilon_\| & 0\\0 & 0 & \varepsilon_\| \end{matrix} \right), \
\begin{array}{lcl}
\varepsilon_\| = \dfrac{\varepsilon_1 d_1 + \varepsilon_2 d_2}{d_1 + d_2},
\\
\varepsilon_\bot = \left(
\dfrac{\varepsilon_1^{-1} d_1 + \varepsilon_2^{-1} d_2}{d_1 + d_2}
\right)^{-1}_.
\end{array}
 \label{eq:diel-tens}
\end{equation}
In the case when $\varepsilon_\|$ and $\varepsilon_\bot$ are of opposite sign, i.e. $\varepsilon_\| \varepsilon_\bot < 0$, a material is called indefinite~\cite{smith_schurig} or hyperbolic~\cite{noginov2009, shalaev2010}. Since hyperbolic metamaterials allow high-$k$ modes they may have very high photonic density of states~\cite{poddubny, sipe2011}. 

However, it should be noticed that the local effective medium model is not accurate and used in the present paper only to show nonlocal behaviour of the multilayered metamaterial by comparison between results provided by the local model and the actual description. The nonlocal effective medium model (see Ref.~\cite{chebykin}) is capable of providing analytic results coinciding with ones of transfer matrix method which is the most correct electromagnetic description of multilayers.

Analysis of MMDN was accomplished in the work~\cite{vinogradov} for lossless metamaterials. However, proper attention was not devoted to the imprescriptible attribute of the structure - nonlocality. Inapplicability of quasi-static approximation, that is effective medium model, wasn't demonstrated as well. We will start below from a demonstration of the fact that the effective medium model can't be applied for accurate description of the multilayered metal-dielectric metamaterials even in the deep subwavelength limit.

\section{Dispersion Diagrams}
\label{sec:dispdiags}

First of all, we study electromagnetic properties of the lossless structure by means of dispersion diagrams and isofrequency contours. Losses will be introduced in Sec.~\ref{sec:losses}. 

\begin{figure}
\centering\includegraphics[width=\textwidth]{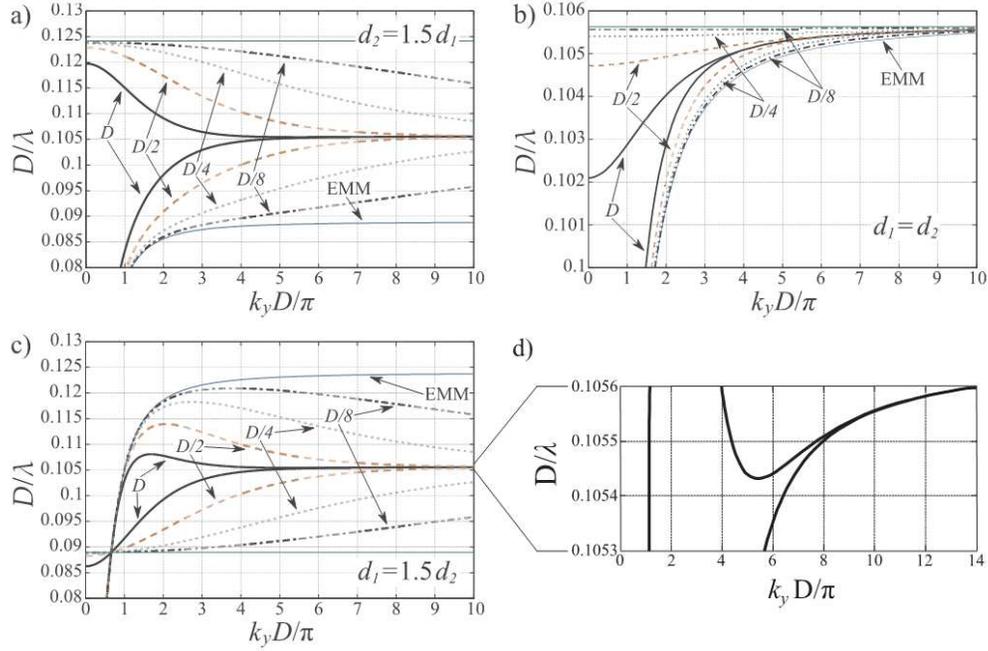}
\caption{Dispersion diagrams for different layers configurations: (a) $d_1 < d_2$, (b) $d_1 = d_2$, and (c) $d_1 > d_2$. Convergence dynamics with the decrease of the period $D$ is shown. (d) Fine structure of the dispersion curves of the metamaterial in the case of thin metal. Four modes are observed to co-exist in the same frequency range. \label{fig:dispdiags}}
\end{figure}

For MMDN with two layers forming its period, using effective medium model and transfer matrix method we obtain two dispersion relations $\omega(\mathbf{k})$, where $\mathbf{k} = (k_x,k_y,0)$ is the wave vector. TM polarization is assumed to take advantage of plasmonic behaviour. The first relation is given by effective medium model:
\begin{equation}
\frac{k_x^2}{\varepsilon_\|} +  \frac{k_y^2}{\varepsilon_\bot} = \left( \frac{\omega}{c} \right)^2,
\label{eq:disp-cryst}
\end{equation}
and another one is given by transfer matrix method, assuming that $k_x^{(i)}$ is the wave number in the $i$-th layer:
\begin{align}
\cos(k_x D) = \cos(k_x^{(1)} d_1)\cos(k_x^{(2)} d_2) - 
\frac{1}{2} \left( \frac{\varepsilon_2 k_x^{(1)}}{\varepsilon_1 k_x^{(2)}} + \frac{\varepsilon_1 k_x^{(2)}}
{\varepsilon_2 k_x^{(1)}} \right) \sin(k_x^{(1)} d_1)\sin(k_x^{(2)} d_2)
\label{eq:disp-matr}.
\end{align}

In this section graphic representation of the dispersion dependencies is chosen in the form of dispersion diagrams for waves travelling along $y$-direction, i.e. along the layers with $k_x = 0$ and presented in Fig.~\ref{fig:dispdiags}. Isofrequency contours will be shown in Sec.~\ref{sec:isofreq}.

General view of the structure is presented in Fig.~\ref{fig:dispdiags}.a. The structure consists of metal and dielectric layers and indefinite in all directions. Period of the structure is $D = d_1 + d_2 = 62.5$nm, where $d_1$ and $d_2$ are the layers thicknesses. Dielectric function for the dielectric layers is constant and equals to $4.6$, while metal layers are described by dielectric function of the Drude form:
\begin{equation}
\varepsilon_2(\omega) = 1 - \frac{\omega_p^2}{\omega (\omega + i\Gamma)},
\end{equation}
with the plasma wavelength $\lambda_p = 2\pi c/\omega_p = 4D$ and the damping coefficient $\Gamma$. The multilayered metamaterial with 3 different ratios $d_1 / d_2$ of layers thicknesses (2/3, 1 and 3/2, respectively) and different values of $D$ were chosen in order to demonstrate different behaviours of the structure. 

The case of the full period equal to $62.5$~nm has been considered in Ref.~\cite{orlov}. Here we also consider structures with periods equal to $D/2, D/4$, and $D/8$ and compare their behaviour with one of the full period. Astonishingly, the behaviour observed for the full period that is 10 times smaller than the wavelength remains qualitatively the same for structures having 2, 4 and 8 times smaller periods (i.e. 20, 40 and 80 times smaller than wavelength), as it is shown in Fig.~\ref{fig:dispdiags}. The convergence to an asymptote corresponding to surface plasmon-polariton (SPP) resonance becomes weaker as the branches reach asymptote at larger values of $k_y$, but a topology of the branches remains the same. A pronounced tendency is such that the decrease of the period pushes one of dispersion curves closer to one predicted by the effective medium model for small $k_y$. For large $k_y$ that convergence to effective medium model is weaker. Another dispersion branch tends to converge to a straight line corresponding to a volume plasmon with the decrease of the period. Volume plasmon is at the frequency where $\varepsilon_{||} = 0$ and it is marked out with green thin line in Fig.~\ref{fig:dispdiags}. Positions of these two lines, effective medium model line and volume plasmon line, depend on the $d_1 / d_2$ ratio. That means that the convergence also depends also on the thicknesses ratio. In the case of equal thicknesses the convergence is faster than for different thicknesses. However, it is clear that even in the quasi-static limit correspondence between the effective medium model predictions and the actual dispersion properties of the metamaterial is not satisfactory, especially for high-$k$ waves with relatively large $k_y$ which are mainly used in subwavelength imaging. We have also discovered from dispersion diagram analysis that the metamaterial in the case of thin metal ($d_1>d_2$) possesses a fine structure shown in Fig.~\ref{fig:dispdiags}.d and supports four modes co-existing in the same frequency range. 

\section{Isofrequency Contours}
\label{sec:isofreq}

\begin{figure}
\centering\includegraphics[width=11cm]{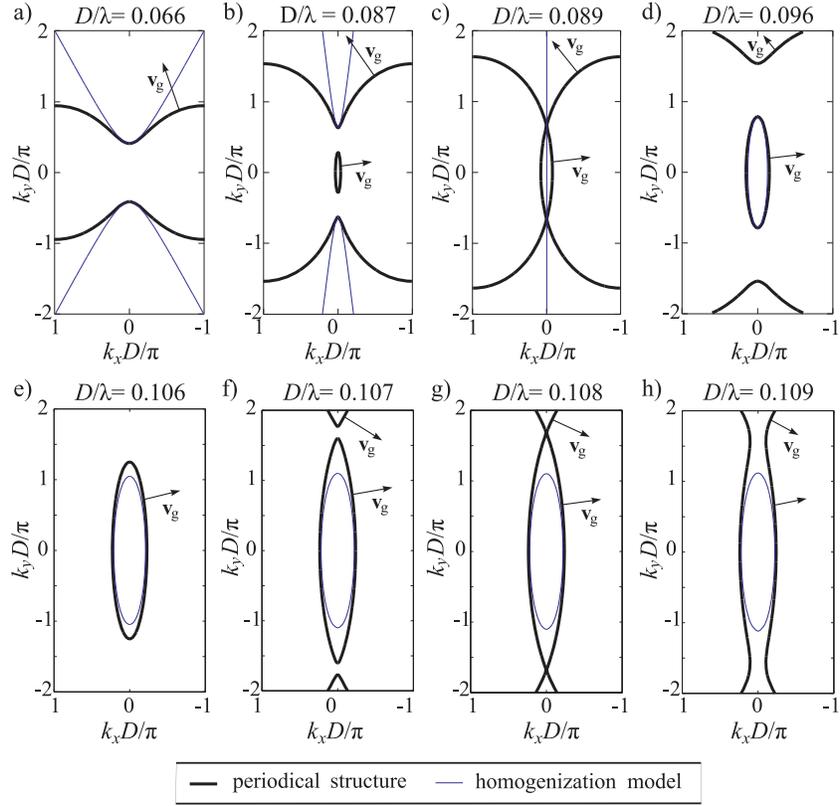}
\caption{Isofrequency contours for the case of $d_1 > d_2$. Group velocities are shown for realistic contours by arrows. \label{fig:cisofreq}}
\end{figure}

\begin{figure} 
\centering\includegraphics[width=11cm]{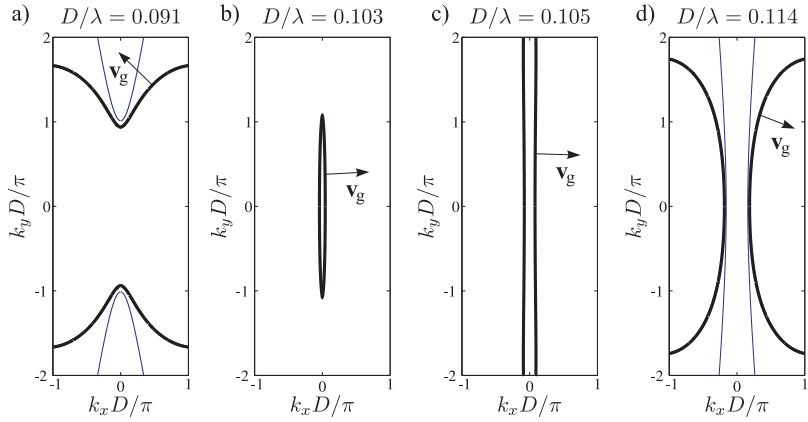}
\caption{Isofrequency contours for the case of $d_1 = d_2$. \label{fig:bisofreq}}
\end{figure}

\begin{figure} 
\centering\includegraphics[width=11cm]{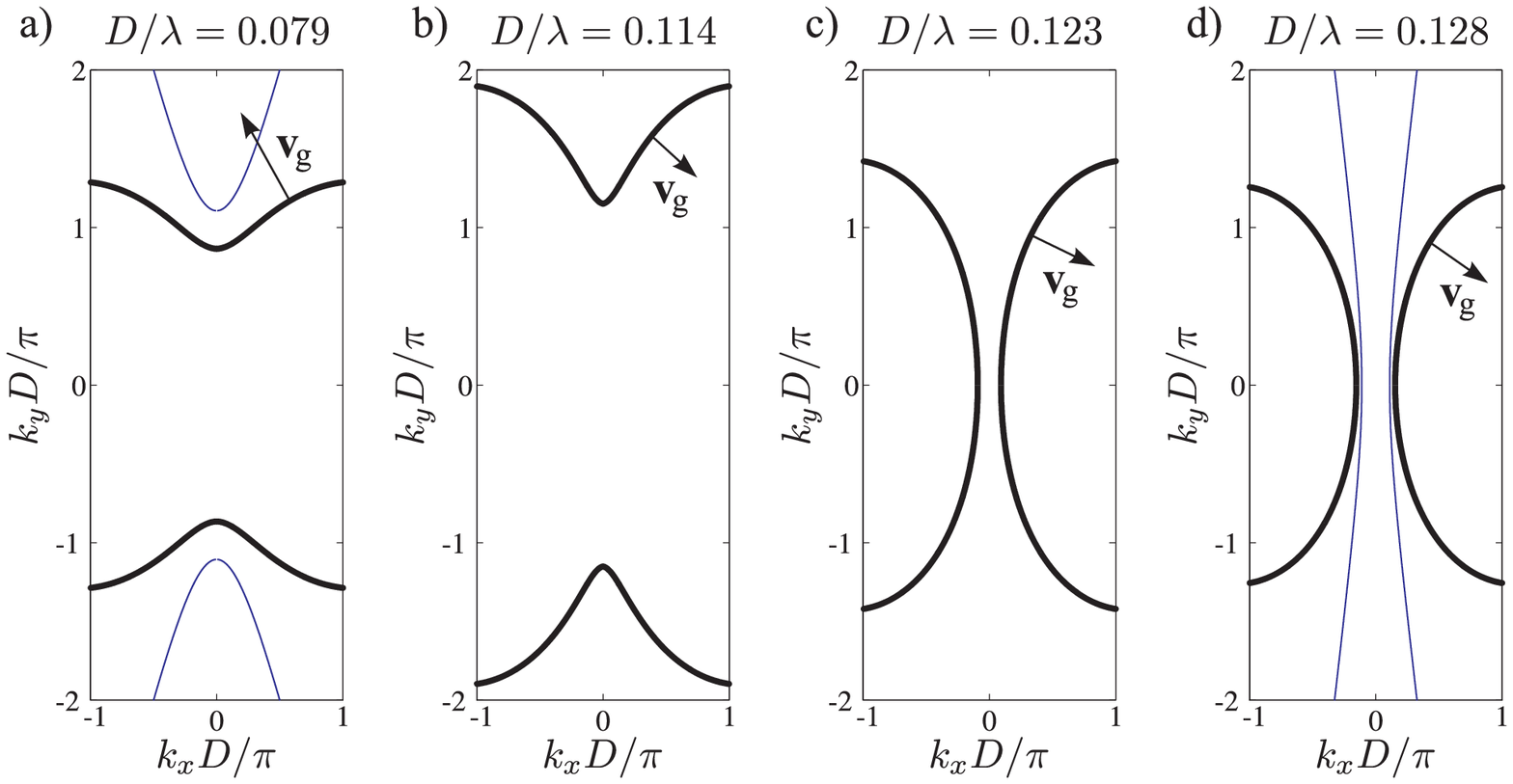}
\caption{Isofrequency contours for the case of $d_1 < d_2$. \label{fig:aisofreq}}
\end{figure}

Another powerful tool to analyze dispersion relation $\omega(\mathbf{k})$ is analysis of isofrequency contours, which are obtained if one fixes the frequency $\omega$ and plots $k_x(k_y)$ dependence. Isofrequency contours for three different thicknesses ratios of multilayered metamaterials at certain frequencies are presented in Figs.~\ref{fig:cisofreq}-\ref{fig:aisofreq}. You can find isofrequency contours for the whole optical frequency range in the video attachment of the present paper.


For the configuration corresponding to the case of thin metal ($d_1>d_2$), which has the most unusual dispersion properties, isofrequency contours are presented in Fig.~\ref{fig:cisofreq} for 8 different frequencies. Examination of the contours reveals that at certain frequencies they have very unusual topology. For $D/\lambda=0.066$ (see Fig.~\ref{fig:cisofreq}.a), two hyperbolic isofrequency contours predicted by effective model are well reproduced by MMDN. Deviations from the effective medium model appear only at large $k_x$ close to the edges of the Brillouin zone, where the homogenization procedure is forbidden since the wavelength in the material becomes comparable to its period. For $D/\lambda=0.087$ (see Fig.~\ref{fig:cisofreq}.b), where the frequency is closer to the SPP resonance, in addition to the hyperbolic contours MMDN demonstrates an elliptic-like contour right around the center of Brillouin zone. This region of the Brillouin zone is usually treated as one where the quasi-static approximation is to be applicable since eigenwaves have small wave numbers, i.e. small spatial variation. However, inside layers fields vary dramatically and this leads to emergence of nonlocality. At slightly higher frequency $D/\lambda=0.089$ (see Fig.~\ref{fig:cisofreq}.c), elliptical and hyperbolic contours observed at lower frequencies are transformed into a pair of crossing ellipses. In this case, the effective medium model predicts completely flat isofrequency contour which is perfect for realization of canalization regime for subwavelength imaging~\cite{Belov,he}, but it is not in agreement with realistic dispersion properties of MMDN. At the frequency $D/\lambda=0.096$ (see Fig.~\ref{fig:cisofreq}.d), an elliptical isofrequency contour predicted by effective medium model is accompanied by two hyperbolic contours, which disappear at higher frequencies, for example, at $D/\lambda=0.106$ (see Fig.~\ref{fig:cisofreq}.e). However, they appear once again at slightly higher frequencies, e.g. $D/\lambda=0.106$ (see Fig.~\ref{fig:cisofreq}.f). Presence of elliptic isofrequency contours is a clear sign of nonlocality in the MMDN. At $D/\lambda=0.108$ (see Fig.~\ref{fig:cisofreq}.g) the topology of isofrequency contours is similar to ones at $D/\lambda=0.089$ (the crossing ellipses), but at slightly higher frequency $D/\lambda=0.109$ (see Fig.~\ref{fig:cisofreq}.h) the contours acquire even more unusual, pitcher-like form, which also has nothing to do with effective medium model predictions.

For the rest of thicknesses ratios when the thicknesses are equal or dielectric is thinner than metal isofrequency contours are presented in Figs.~\ref{fig:bisofreq},~\ref{fig:aisofreq}. There are not new topological structures that are absent in the case of thin metal, however, hyperbolic-like and elliptical contours are still there. One should note flat isofrequency contours in Fig.~\ref{fig:bisofreq}c that appear at the frequency where $\varepsilon_\bot = \infty$. It is known that flat isofrequency contours are in use for subwavelength image canalization~\cite{he,Belov2005}. Also, in the case of equal thicknessess plotted in Fig.~\ref{fig:bisofreq} we see that isofrequency contours of effective medium model are in a quite good agreement with realistic ones.

\section{Complex Band Structure}
\label{sec:losses}

Dispersion relation given by Eq.~\ref{eq:disp-matr} allows us to study complex modes of the structure immediately complex $k_y$ is considered. There are plenty of methods complex dispersion equation can be solved by, including a method based on the argument principle theorem~\cite{findzeros}, operator theory methods~\cite{shanhuifan}, and numerical FEM methods~\cite{shvets2007,shvets2011}. However, in order to find roots of Eq.~\ref{eq:disp-matr} in the complex plane we employ a combination of change of sign method and Newton's one. 

\begin{figure} 
\centering\includegraphics[width=12cm]{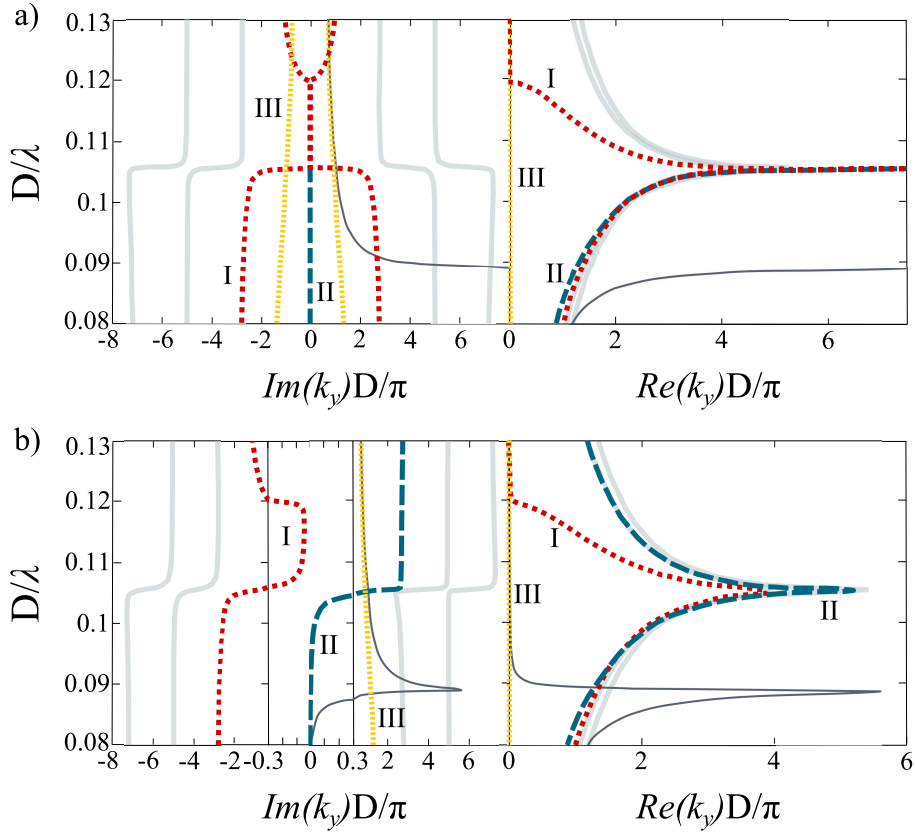}
\caption{Complex band structure of the structure with $d_1<d_2$ in the absence of losses (a) and with the losses (b). Imaginary parts of the modes are on the left, real parts are on the right. Low-loss modes are marked with colors, gray curves corresponds to high degenerative modes. \label{fig:losses_d215d1}}
\end{figure}

\begin{figure} 
\centering\includegraphics[width=12cm]{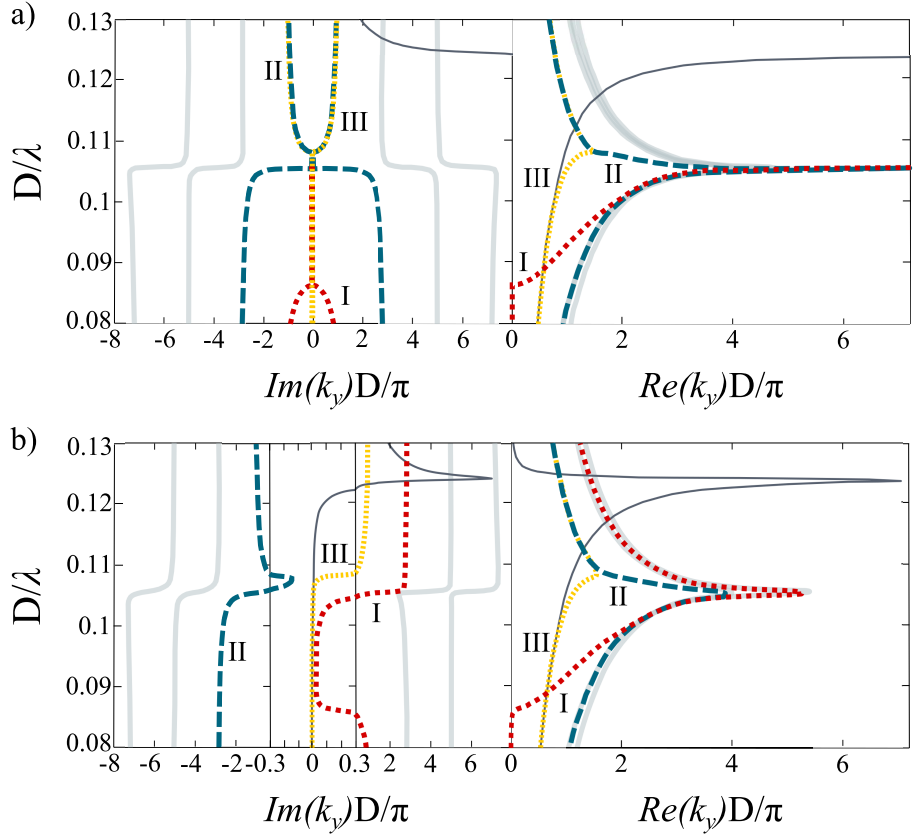}
\caption{Complex band structure of the structure with $d_1>d_2$ in the absence of losses (a) and with the losses (b). Imaginary parts of the modes are on the left, real parts are on the right. \label{fig:losses_d115d2}}
\end{figure}

In Figs.~\ref{fig:losses_d215d1},~\ref{fig:losses_d115d2} dispersion diagrams of lossless ($\Gamma\rightarrow0$) and lossy structures showing complex modes are presented. We show positive real part of the wave vector with corresponding imaginary part. Due to the periodicity and unboundedness of the structure its complex eigenvalue spectrum contains an infinite number of modes. However, only modes having low imaginary parts of its wavenumber are of interest since they can travel significant distances. For example, modes having $Im(k_y)<\pi/D$ are able to propagate to distances more than $D/2\pi$. We restrict imaginary part of $k_y$ by the value of $8\pi/D$ in our calculations. 

In the absence of losses the structure supports the set of waves having the wavevectors $\pm k$ and $\pm k^*$. Thus, there are forward and backward waves along with their complex conjugations, and imaginary parts of the modes are symmetric in Figs.~\ref{fig:losses_d215d1}a,~\ref{fig:losses_d115d2}a, . It follows from the fact that if dielectric permittivities of layers are real then the right part of Eq.\ref{eq:disp-matr} is a holomorphic function whose restriction to the real numbers is real-valued. 

The most important three modes are marked in Figs.~\ref{fig:losses_d215d1},~\ref{fig:losses_d115d2} with different styles and colors while high-order complex modes are in gray color. Effective medium model results are presented in the figures as well (thin curves). Modes showed in Fig.~\ref{fig:dispdiags} of Sec.~\ref{sec:dispdiags} can be recognized by their zero imaginary part in regions of propagation. Where in Fig.~\ref{fig:dispdiags} there were band gaps, here we have fully complex modes. At the SPP resonance frequency real part of the propagating mode goes to infinity, while imaginary part degenerates at this point what can be seen on the example of mode II in the case of $d_1<d_2$ and on the mode I in the case of $d_1>d_2$. One can also observe an effect of decay switching in high-order modes: after a mode undergoes the SPP resonance its imaginary part jumps from one branch to another. In the case of $d_1 < d_2$ that is presented in Fig.~\ref{fig:losses_d215d1}a there is a very special mode III that is fully imaginary in the presented frequency range, and its real part is dispersionless and equals to zero. This feature is very unique since that mode is not sensitive to the resonance. Its presence is highly related to effective medium model as the latter has the resonance at another frequency and at higher frequencies converges to the entirely imaginary mode III while at low frequencies it converges to the mode II.

\begin{figure}
\centering\includegraphics[width=9cm]{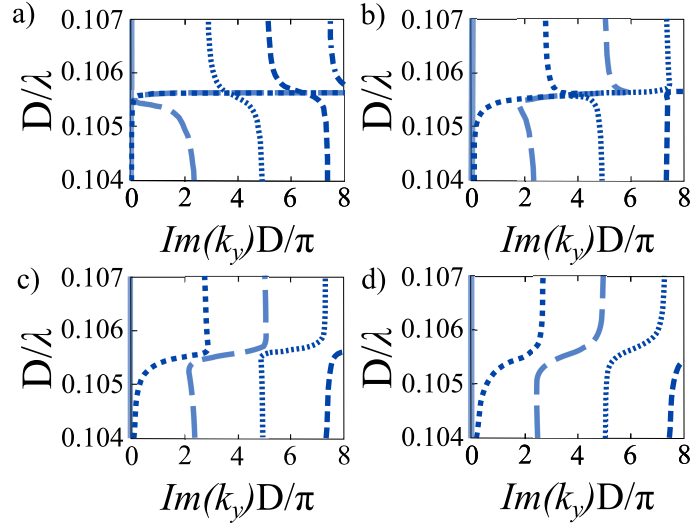}
\caption{Mode transformation mechanism in MMDN: a) $\Gamma = 1.885 \cdot 10^{11} s^{-1}$, b) $\Gamma = 5.655 \cdot 10^{12} s^{-1}$, c) $\Gamma = 9.425 \cdot 10^{12} s^{-1}$, and d) $\Gamma = 1.508 \cdot 10^{13} s^{-1}$ \label{fig:lswitch}}
\end{figure} 

However, zero losses aren't realistic and unachievable in practice. Thus, we introduce attenuation in metal to study effect of losses in metal layers on electromagnetic properties of MMDN. Increasing losses smoothly we find that it changes step-like behaviour of the modes at the SPP resonance. That is, a specific form of these steps is determined by the value of losses in MMDN. With change of $\Gamma$ modes go through a number of transformations~\cite{kivshar_2011}. Mechanism of modes transformation is exposed in Fig.~\ref{fig:lswitch}. When losses tend to zero, at the resonance positive imaginary branches jump to their left neighbours with growth of the frequency, while negative branches jump to their right neighbours. Then, increase of the losses level forces branches to modify their topology and after a number of metamorphoses at the resonance both positive and negative imaginary branches jump to their right neighbours with growth of the frequency.

This way we come to realistic level of losses. $\Gamma = 1.734 \cdot 10^{13} s^{-1}$ is supposed in our calculations. Since now we analyze complex modes in the case of realistic losses comparing complex band structure with the lossless case to show how losses affect the behaviour of the structure. In the presence of losses, as soon as the damping coefficient $\Gamma$ in the dielectric function of metal $\varepsilon_2$ becomes positive non-zero and $\epsilon_2$ becomes a complex number, the right part of Eq.~\ref{eq:disp-matr} can be complex-valued with real $k_y$. Consequently, the modes confluence is removed and the conjugated waves with $\pm k^*$ are no longer supported by the structure. However, correspondence between positive and negative real and imaginary parts can be quite diverse because of the absence of symmetry in the imaginary parts of eigemodes (see Figs.~\ref{fig:losses_d115d2}b,~\ref{fig:losses_d215d1}b). 

In Figs.~\ref{fig:losses_d115d2}b,~\ref{fig:losses_d215d1}b we still have decay switching effect that is seen better in high degenerative modes. It's expected for modes we have observed in Figs.~\ref{fig:losses_d115d2}a,~\ref{fig:losses_d215d1}a: losses of a mode are grown up dramatically when it enters into a band gap, but for complex modes such behaviour is non-obvious. Also, as soon as one moves from the real axis to the complex plane every mode becomes complex one and it becomes hard to distinguish between propagating and evanescent modes. The case of $d_1 > d_2$ is shown in Fig.~\ref{fig:losses_d215d1}b. Just that very case corresponds to simultaneous presence of forward and backward modes in the same frequency range ($D/\lambda: 0.105 - 0.108 $) for propagation along the layers. That is, the beam splitting phenomenon revealed in~\cite{orlovPRB} remains even after introduction of realistic level of losses. Decay switching effects are observed there for high degenerative modes and the modes I, III while the mode II has narrow propagation domain with high losses in the rest of it.  

Finally, it's noteworthy that effective medium model predicts resonance behaviour for imaginary part of $k_y$. In reality, however, the structure is functioning in a regime with infinite number of modes which imaginary parts have step-like behaviour at the resonance. 

\section{Field Profiles}
\label{sec:profiles}

\begin{figure}
\centering\includegraphics[width=0.9\textwidth]{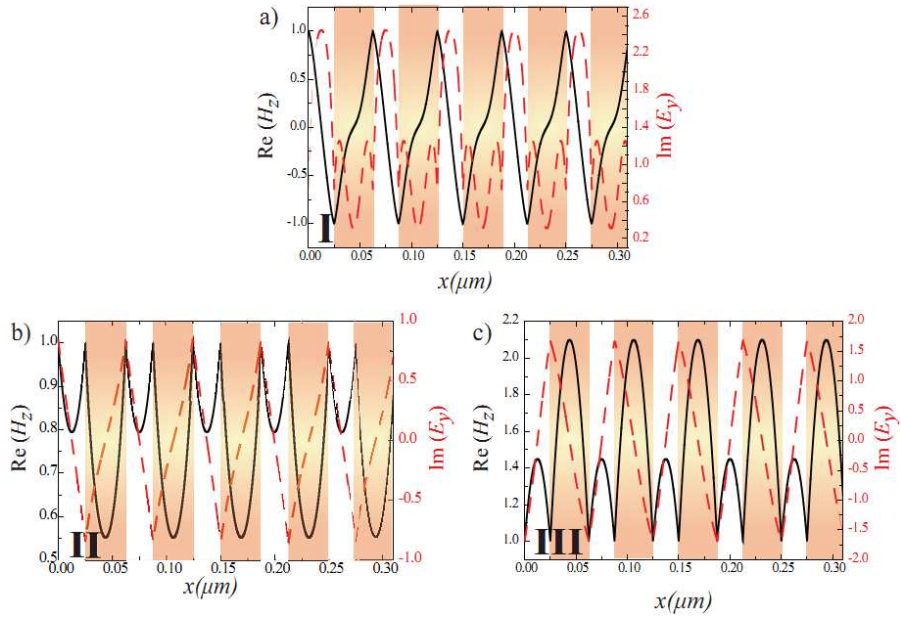}
\caption{Field profiles in the case of $d_1<d_2$ for three modes at the frequency of $D/\lambda=0.0875$. Solid curves correspond to the real part of the magnetic field, dashed curves correspond to the imaginary part of the electric field. \label{fig:profiles_thickmetal}}
\end{figure} 
\begin{figure}
\centering\includegraphics[width=0.9\textwidth]{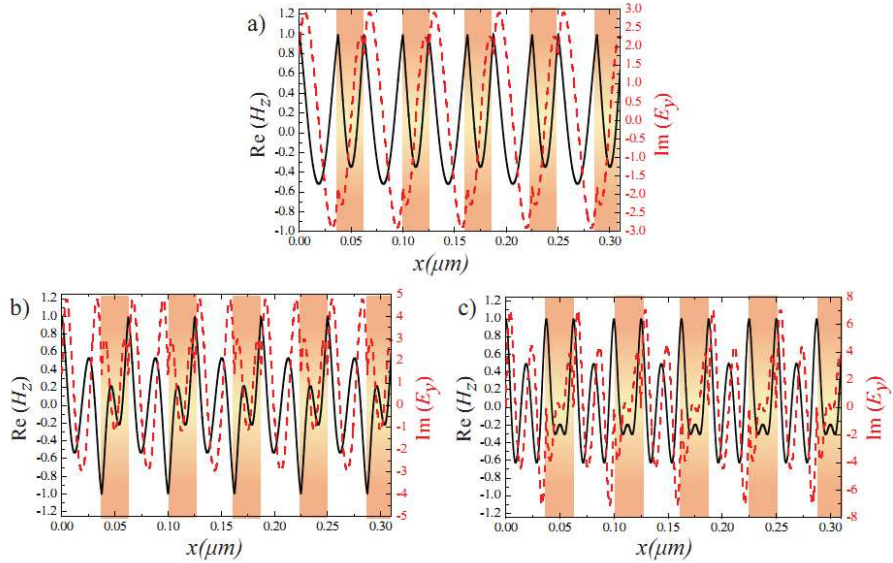}
\caption{Field profiles of high-order modes corresponding to dispersion curves shown in Fig.~\ref{fig:lswitch}.d. \label{fig:profiles_lswitch}}
\end{figure}

In order to obtain deeper understanding of the nature of the studied complex eigenmodes of the system, we have plotted profiles of the magnetic and electric fields of these modes. Fig.~\ref{fig:profiles_thickmetal} shows the profiles of the 3 branches for the case of the thicker metal at the frequency $D/\lambda=0.0875$. We first note that the magnetic field of the mode $\mathrm{I}$ averaged over each layer is equal to zero. 
It means that the mode I is effectively longitudinal. Such  modes are well known in plasma physics as \textit{Langmuir} modes~\cite{andreev}. It has been also recently shown that such exotic modes may exist in the waveguides with metal cladding and core made of hyperbolic media~\cite{Bogdanov}. We can see however, that the longitudinal nature of the mode holds only within the effective media approximation and the local magnetic field inside each layer does not vanish. Modes $\mathrm{II}$ and $\mathrm{III}$ are conventional transverse modes. 

\begin{figure}
\centering\includegraphics[width=0.9\textwidth]{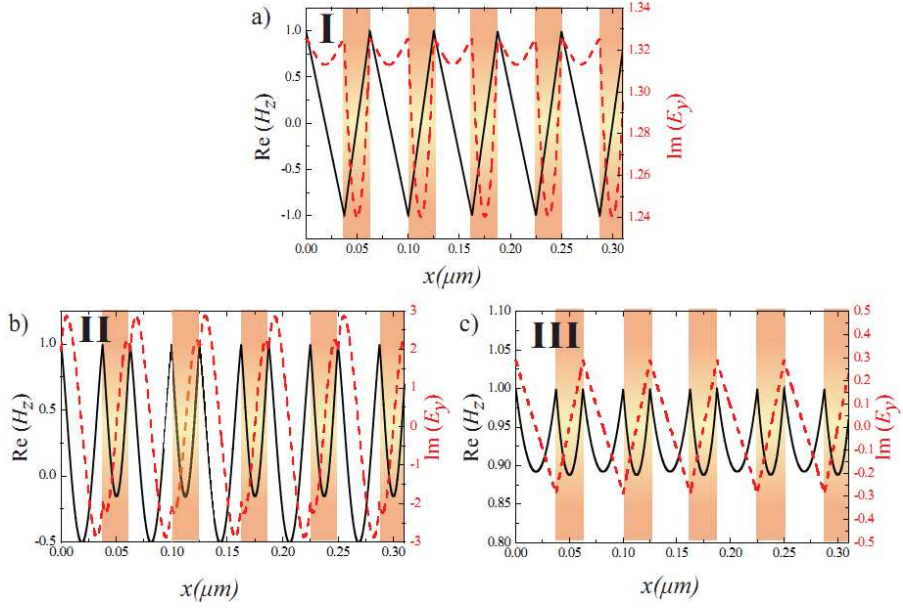}
\caption{Field profiles in the case of $d_1>d_2$ for three modes at the frequency of $D/\lambda=0.0875$. Solid curves correspond to the real part of the magnetic field, dashed curves correspond to the imaginary part of the electric field. \label{fig:profiles_thinmetal}}
\end{figure} 

Another interesting feature that can be noted on the profiles is the complicated field distribution of the electric field inside the metal layer for mode $\mathrm{I}$. Conventionally the modes existing in metal-dielectric structures are coupled surface plasmon resonances at the individual metal-dielectric interfaces. Thus, there may exist no more than one extremum of the field distribution function in each layer. This condition however does not hold for the case of the complex modes. Indeed, if we consider the modes with real and imaginary part of the propagation constant, the expression for the transverse component of the wavevector in each layer reads:
\begin{align}
k_x=\sqrt{\varepsilon_ik_0^2-k_y^2}=\sqrt{\varepsilon_ik_0^2-\mathrm{Re}(k_y)^2+\mathrm{Im}(k_y)^2-2i\mathrm{Re}(k_y)\mathrm{Im}(k_y)}.
\end{align}
It is clear that when $\mathrm{Im}(k_y)$ is large enough, the $k_x$ gains large real part and the mode starts to propagate inside the structure. This leads to nonzero values of the transverse component of the Poynting vector inside the layers and to the complex shape of the field distribution inside the layers.
To illustrate how the imaginary part of $k_y$ affects the shape of the field inside the layers we have plotted the profiles of the high order complex modes [see Fig.~\ref{fig:profiles_lswitch}]. 
We can see that while we are switching to higher order modes, having the larger imaginary part of $k_y$ the field profile shape becomes more and more complicated. These modes can be regarded as a specific type of coupled waveguide modes. The waveguide mode condition can be roughly estimated as $k_{i,x}d_i=\pi n$, where $i=1,2$ and $n$ - is an integer. Thus, higher switching between the complex modes can be regarder as switching between different waveguide modes in the structure.

We have also plotted the profiles of the three mode branches for the thinner metal case [see Fig.~\ref{fig:profiles_thinmetal}]. We can see that in this case we also have one longitudinal mode $\mathrm{I}$ and two transverse modes. In this case however the mode $\mathrm{II}$ has the complicated structure of the electric field as the one defined by the larger values of $\mathrm{Im}(k_y)$.


\section{Conclusion}
Using dispersion diagrams along with isofrequency contours we have analyzed different characteristic configuration of MMDN both in the absence and in the presence of losses. Study of real-valued eigenmodes has revealed the impossibility of its description in terms of the local effective medium model even in the deep subwavelength limit. With the introduction of realistic losses, complex band structure of MMDN was revealed. Generally, nonlocality in MMDN is slightly affected by losses. Thus, in real structures with losses variety of spatial dispersion effects is expected.

\section*{Acknowledgements}
This work has been supported by the Ministry of Education and Science of Russian Federation, the Dynasty Foundation, 
EPSRC (UK), and the Australian Research Council.

\end{document}